\documentstyle[prl,aps,twocolumn,floats]{revtex}
\begin{document}
\title
{Comments on the paper ``Bare Quark Matter Surfaces of Strange Stars and
$e^+ e^-$ Emission''}



\maketitle

We would like to point out here that although the idea expressed in the
above mentioned paper\cite{1} that the bare surface of a hot strange star
may be a strong source of $e^+ e^-$ pairs of luminosity $L_{e^\pm} \sim 3
\times 10^{51}$ergs s$^{-1}$ for a duration of $\sim 10$ s
appears to be novel and exciting, it is, unfortunately, unlikely to work
because of several reasons.
It is true that, it has already been hypothesesized by previous
authors\cite{2} that the bare surface of a strange star may be endowded
with a superstrong electrostatic field of magnitude $E \sim 5\times
10^{17}$V cm$^{-1} \sim 1.5 \times 10^{13}$ statvolt cm$^{-1}$. Although
such estimates, in general, could be fairly uncertain, let us momentarily
assume that there is indeed such a field on the surface of the strange
star  over a height of $\Delta R\sim 10^{-10}$cm. As correctly envisaged
in ref. 1, such a strong value of $E >>E_{cr}= 1.3 \times 10^{16}$V
cm$^{-1}$ is likely to convert the virtual pairs of the QED vacuum into
real ones, i.e, create real $e^+ e^-$ pairs. Let us consider the region of
the electric field as a parallel plate capacitor.  The maximum value of
the total energy liberated by the discharge of the capacitor, in the form
of pairs and photons,  would be the stored electrostatic energy of the
capacitor :
\begin{equation}
W_s = {E^2 \over 8 \pi} (4 \pi R^2) \Delta R \sim 10^{28} ~{\rm ergs}
\end{equation}
where we have taken $R \sim 10^6$cm. In fact, in realistic cases, the
system would adjust itself in such a manner that, atleast in a steady
state situation, the value of $E$ would be well below $E_{cr}$ even if is
assumed that momentarily it shoots up above $E_{cr}$. It is clear that
this process is driven by electron thermal energy energy. Once the
supposed initial discharge takes place the electrons cool and become
completely degenerate. Now the pair production process is is halted due to
complete blocking of the final state.
 It may be argued
that once the capacitor is discharged by emitting an energy $W_s
<10^{28}$ergs, the left over electrons in the body/core of the star may
reorganize themselves to resurrect the capacitor. But, then, it should be
borne in mind that  the proposed pair generation mechanism does not
harness energy from the quantum vacuum; the source of energy which drives
Usov's mechanism is {\em the transport of trapped thermal energy} from a
supposed young and hot stange stellar core. This might be possible
(although we feel that the actual value of $E$ would always be lower than
$E_{cr}$ if energy is supplied by the hot core by means of electron
conduction or other electromagnetic means because it can be shown that,
contrary to Usov's claim, energy transport by bulk convection is highly
inappropriate here.  Irrespective of the actual electron-reorganization
time scale, or even {\em irrespective of whether pair-proction discharge
leads to a quenching of the suface electric field or not} we can crudely
estimate time needed to release the trapped thermal energy  from the core
by ignoring the fractional charge of the the quarks and crudely treating
the medium as a baryonic plasma. Given a density of $\rho \sim 4\times
10^{14}$g cm$^{-3}$, a value of relative electron fraction $Y_e
\sim 10^{-3}$, a typical $e^\pm \gamma$ interaction cross-section
$\sigma_{\rm Thomson} \sim 6.65
\times 10^{-25}$cm$^2$, the typical mean free path of the
electrons/photons would be $\lambda_e \sim (n Y_e \sigma_{\rm
Thomson})^{-1} \sim 10^{-11}~ {\rm cm}$ where $n$ is the number density of
the baryons.  The corresponding time scale for diffusive energy transport
from the core would be $t_e \sim 6 R^2 / (c \lambda_e) \sim 10^{13} ~{\rm
s}$. In the adjoining reply, Usov has {\em confused  the parameters}
$R\sim 10^6$cm with $\Delta R\sim 10^{-10}$cm  while estimating $t_c$.
When one actually uses $R$ for finding the core-surface temperature
gradient, it follows that, $t_c$ would be many orders of magnitude higher
than $\sim 10$s.

Thus, clearly, the {\em energy transport from the hot core can take place
on time scales sec or less only via the emission of neutrinos} or other
suitable particles having sufficiently weak interaction
cross-section\cite{3,4}.  The probable URCA type reactions by which the
strange star may cool could be $u+e^- \rightarrow d+\nu_e$, $u+e^-
\rightarrow s+ \nu_e$, $d \rightarrow u+e^- +{\bar \nu_e}$ and
$s\rightarrow u +e^- +{\bar \nu_e}$\cite{2}.  It may be worthwhile to
recall here even if we conceive of a pure $e^+ e^- \gamma$ fireball of
energy content $\sim 10^{51}$erg, confined within a radius of $\sim
10^7$cm, say, by a super strong magnetic field $\sim 10^{16}$G, the
photons or pairs will be locked together by the excessive high
electromagnetic opacities, and the only way this pure electromagnetic
fireball can cool is via emission of neutrino pairs: $e^+e^- \rightarrow
\nu_e +{\bar \nu_e}$, $\gamma +e^- \rightarrow e^- +
\nu_e +{\bar \nu_e}$.
 To conclude, Usov obtained this picture  by {\em tacitly assuming that
even when the final state is blocked in the absence of any viable
electromagnetic energy transport on a time scale of $\sim 10$s,  the pair
production rate remains unabated}.

\vskip 0.5cm
Abhas Mitra,  amitra@apsara.barc.ernet.in \\
Theoretical Physics Division\\
BARC, Mumbai-400085, India\\

\end{document}